\documentclass[amsmath,amssymb]{revtex4}
\usepackage{graphicx}
\usepackage{dcolumn}
\usepackage{bm}

\begin{document}

\title{Contraction gradient induced microcracking in hardened cement paste} \author{Jan Bisschop, Falk K. Wittel}
\affiliation{ETH-Zurich, Institute for Building Materials, Dept. of Civil Engineering, CH-8093 Zurich, Switzerland}

\begin{abstract}
Drying induced cracking of concrete surfaces and repair layers is a common problem. A principal cause for this type of cracking is the moisture and resulting contraction gradient that develops in the cement paste matrix upon drying. This phenomenon has been experimentally quantified in unconfined hardened cement paste samples using a fluorescent resin impregnation technique. The effects of sample thickness and drying method on surface crack density and crack penetration depth are reported and explained. Finite element modelling of moisture gradients indicate the important role of the film coefficient in desiccation cracking of unconfined samples. The critical thickness for samples to remain crack-free upon drying was in the range of 2-5 mm depending on drying method. In thicker samples a crack spacing doubling process was observed that is in agreement with theoretical predictions.
\end{abstract}

\maketitle

\section{Introduction}\label{sec-intro}
Polygonal fracture patterns due to desiccation (drying) shrinkage occur commonly in natural and man-made materials such as clay, wood, paint and cement-based materials. Desiccation cracking has been well studied in materials under external confinement, such as suspension systems on a restraining substrate (e.g., [1-4]) or concrete in restrained shrinkage tests (e.g., [5-8]). In this paper we study unconfined hardened cement paste samples that crack upon drying solely due to the development of moisture/contraction gradients. It is well known that externally unconfined cement paste cracks upon drying [9-11], but very few quantitative experimental data exist for this particular case. An understanding of how sample thickness and environmental parameters affect surface microcrack densities in cement paste helps to predict the degree of microcracking of drying concrete surfaces such as pavements and bridge decks. Surface (micro)cracks accelerate the ingress of chloride ions that may cause corrosion of the reinforcement bars
[12,13]. Pre-existing surface (micro)cracks may also accelerate deterioration of concrete surfaces by traffic loading [14,15] or freeze/thaw and salt scaling [16].

There are a number of characteristics associated with desiccation cracking in unconfined solids: (i) Since cracks are caused by a moisture/contraction gradient there must be a critical sample thickness and a critical drying rate below which the gradient is too small to cause cracking [10]. (ii) Desiccation cracks in unconfined solids reach a specific maximum depth ($Z_{max}$) relative to the specimen thickness [9,11,17-19] (Fig. 1). (iii) Theoretically models predict that the evolution of a global contraction gradient will lead to a crack spacing and crack depth doubling process in unconfined or 'half-space' specimens [19-25]. A system of parallel cracks forms perpendicular to the free surface with a specific initial spacing (Fig. 1). On the drying surface these cracks are not actually parallel but form cellular patterns. The deeper the cracks grow into the material, the more they interact. At some critical point only
every second crack continues to grow since intermediate cracks become 'shielded' by their neighbours. This doubling process may repeat itself a number of times depending on the system specifics and will lead to a crack depth distribution diagram with two or more peaks.
\begin{figure}[htb]
  \centering{ \includegraphics[width=8.cm]{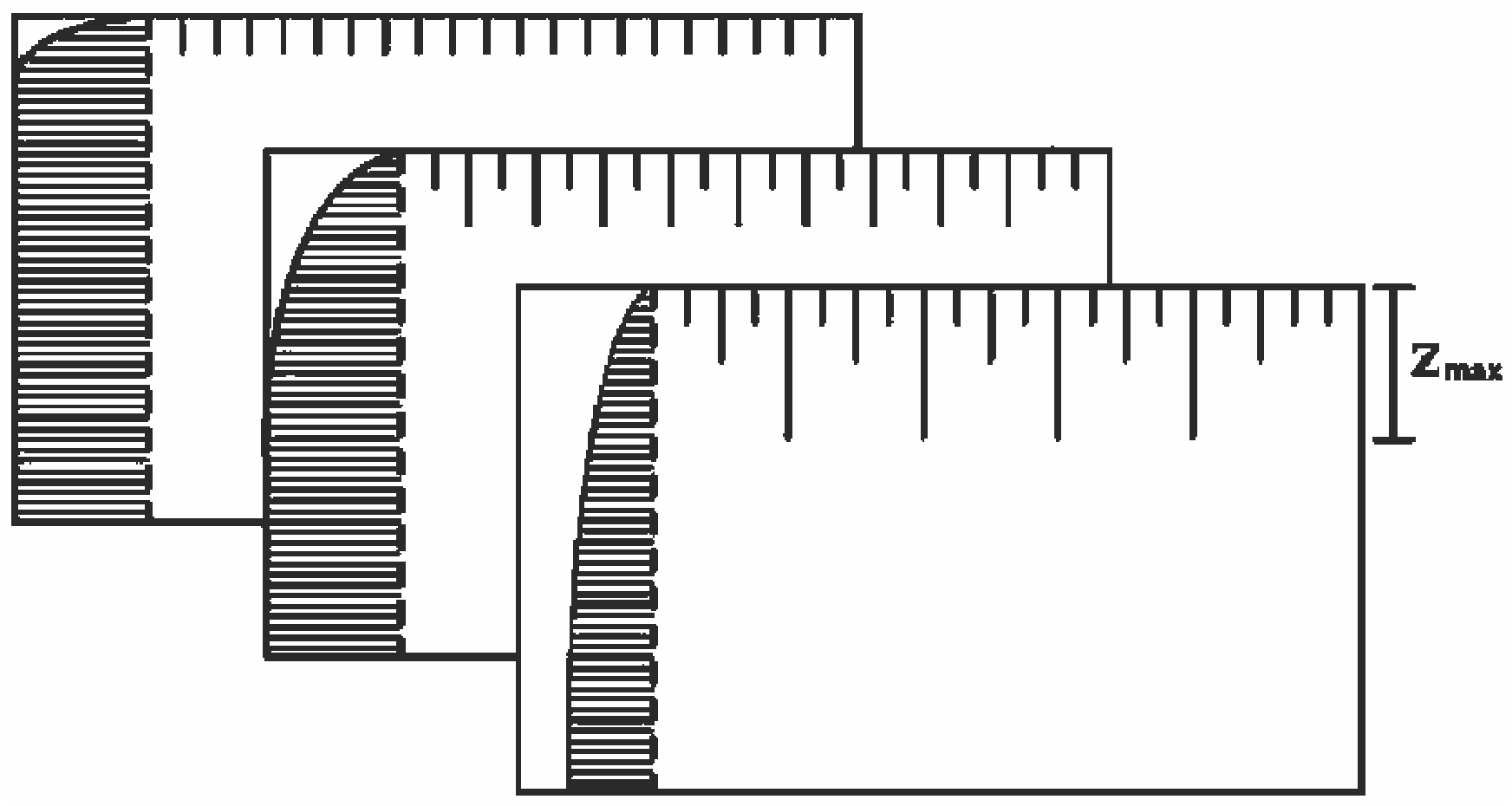} }
  \caption{Model evolution of moisture distribution (shaded grey) and desiccation crack-pattern in unconfined samples.}
\end{figure}

The mathematics of the crack-doubling theories is similar for the cases of drying and cooling [21]. In the first case moisture transport and drying shrinkage gradients control the process, while in the latter one it is governed by heat transport and thermal contraction gradients. Crack spacing doubling has been experimentally observed in the case of thermal shock in certain materials [26,27]. In this paper we investigate if crack spacing doubling also occurs in the case of drying of unconfined cement paste samples. Plain hardened cement paste was studied to avoid the effect of aggregates on crack propagation or even the initiation of cracks by aggregate-restraint that may occur in cementitious composites such as concrete [28,29]. The desiccation crack-patterns were recorded by using a fluorescent resin impregnation technique [28]. The following experimental variables were studied: (i) sample thickness, (ii) drying rate; and (iii) single-sided vs. double-sided drying.

In addition to studying the fundamental aspects of desiccation cracking in unconfined samples, a more practical motivation of this study was to find out how to avoid or reduce desiccation cracking during drying of small samples of hardened cement paste. Dried, crack-free samples are needed for measuring the effect of moisture content or relative humidity on mechanical properties, such as strength, elasticity, creep and pure drying shrinkage of hardened cement paste. This paper reports critical crack-free sample thicknesses for different drying conditions.
\section{Methods}
\subsection{Sample preparation}
All samples consisted of ordinary Portland cement (CEM I 52.5R) with an \textit{added} water-cement weight ratio of 0.5. The cement pastes were thoroughly mixed and cast into steel moulds of 30 x 30 x 15 mm. A total of 200 samples were prepared from 15 cement paste batches. The filled moulds were vibrated for 1 min to remove mixing air-bubbles. The moulds were sealed and stored for 24 h at 20 $^\circ$C. At an age of 24 h, the samples were taken out of the moulds, their weights measured, and placed in an aqueous solution for 27-29 days. The aqueous solution was a calcium saturated solution to prevent Ca(OH)$_2$-leaching from the samples. The average sample moisture absorption during 27-29 days wet-curing was 0.033 g/cm$^3$, which was 80\% of the ultimate absorption measured after 4 months.

After 27-29 days, samples were carefully ground to thicknesses in the range 1-14 mm (see Table 1) on grit-120 grinding paper in a Ca-saturated brine. Due to compaction of the pastes before solidification, a small density gradient existed in the sample. Thinning of the samples was always done by grinding the sample side that was the top surface in the mould (i.e., the low density side). The sample surfaces were finished by fine-grinding on grit-1000, grit-2400, and grit-4000 grinding paper. The samples were 0-30 $\mu$m thinner than the thicknesses shown in Table 1, and the thickness variation within single samples was $\pm 10\mu$m.

Average effective w/c-ratio's of 0.417 and 0.434 for the 2 mm and 14 mm thick samples, respectively, were calculated on basis of a cement density of 3.12 g/cm$^3$, sample volume, and sample weight minus water absorption. At the top of 2 mm and 14 mm samples the w/c-ratio was on average 0.7\% and 4.8\% higher, respectively, than at the bottom. The average Young's modulus of the saturated cement paste was measured to be 12.5 GPa at the age of 28 days.
\subsection{Drying experiments}
Different drying methods were used to induce desiccation cracks (Table 1). Samples were dried either from a single side or from two opposed sides. In order to create these single-sided (\textit{ss}) or double-sided (\textit{ds}) drying samples, the samples were first taken out of the water and surface dried in a conditioning room at 95\% relative humidity (RH). Subsequently, all sides except the drying surface(s) were sealed by two layers of adhesive tape. A climate chamber (Voetsch VC4060) with circulating air at 26 $\pm$ 3\% RH and 20 $\pm$ 0.2 $^\circ$C was used to dry samples. The RH dropped below 30\% within 5 min after placing samples in the chamber and closing the door. The temporal and spatial precision of RH control in the chamber was 1-3\%. The average and standard deviation of the measured RH during all experiments over periods of months was 26\% and 1.3\%, respectively. These severe drying conditions were used for measuring minimum critical crack-free sample thicknesses that are valid for drying in the range between 25\% and 100\% RH. The average equilibrium (ultimate) shrinkage strain of the samples at 26\% RH was 5 mm/m and was measured over the sample thickness.

Series of samples were exposed immediately to the flowing air in the climate chamber and underwent shock drying (series \textit{shock-ss} and \textit{shock-ds}). Other series of samples were placed for 5 h in cylinders (h x $\Theta$ = 10 x 4.5 cm, with a 1.8 cm diameter top hole) inside the climate chamber to create stagnant air above these samples (series \textit{non-shock-ss} and \textit{non-shock-ds}). These samples were placed in the cylinders in the 95\% RH room and were covered with lids until they were put in the climate chamber. The purpose of the latter experiment was to reduce the initial drying rate and avoid shock drying of these samples. After 5 h the cylinders were removed and the samples experienced the same drying condition as the shock drying samples. All single-sided samples dried from their bottom surfaces (i.e., the high density side).

A final series of samples was dried in the chamber of an Environmental Scanning Electron Microscope (ESEM) at 25\% RH and 20 $\pm$ 0.2 $^\circ$C for 5 h, enabling in situ monitoring of the evolution of was held constant at 20 $\pm$ 0.2 $^\circ$C using a Peltier element. The water desiccation crack widths. The sample top surface temperature vapour pressures inside the chamber was set to 4.4 Torr (=590 Pa) to create 25\% RH just above or at the sample surface. This experiment allowed to make a comparison between samples drying in ambient air (at $\approx 10^5$ Pa) and at low pressure (590 Pa). Note that during so-called purging of the ESEM-chamber, i.e., replacing air by water vapour, the samples were very briefly exposed to a RH < 5\%.
\begin{table*}[htb]\label{tab1}
\caption{Overview of experiments.}
  \begin{tabular}{p{3cm}p{5cm}p{3cm}p{3cm}}
    \textbf{drying method}& \textbf{'far-field' drying conditions }& \textbf{drying } & \textbf{sample thickness}\\
     &  & \textbf{durations }& \textbf{[mm]}\\ \hline\hline
    \textit{shock-ss} & 26\%RH, 20$^\circ$C, ambient P & 5h, 2 months & 3, 4, 5, 6, 8, 11, 14\\
    \textit{shock-ds} & 26\%RH, 20$^\circ$C, ambient P & 5h, 2 months & 2, 3, 4, 6, 8, 14\\
   \textit{non-shock-ss} & 26\%RH, 20$^\circ$C, ambient P & 5h, 2 months & 5, 8, 11, 14\\
   \textit{non-shock-ds} & 26\%RH, 20$^\circ$C, ambient P & 5h, 2 months & 5, 8, 11, 14\\
   \textit{shock-ESEM} & $\approx$25\%RH, $\approx$20$^\circ$C, 590 Pa & 5h & 1, 2, 3, 4, 6\\
  \end{tabular}
\end{table*}
\subsection{Crack pattern analysis}
After 5 h or 2 months drying, the samples were impregnated with epoxy resin containing a fluorescent dye. The resin was poured over the drying surfaces and penetrated the cracks by capillary suction. Excess hardened resin was pealed off from the drying surface and the impregnated top layer ($\approx 100 \mu$m) was removed by grinding on grit-1000 sand paper to make the cracks clearly stand out. Cracks with a depth of less than 100 $\mu$m therefore remained undetected in this study. The minimum impregnatable crack width lays below 0.5 $\mu$m. The crack-patterns on the
drying surfaces were photographed in a dark room with fluorescent light and long exposure times using an 8 megapixel digital camera.
\begin{figure}[htb]
  \centering{ \includegraphics[width=8.cm]{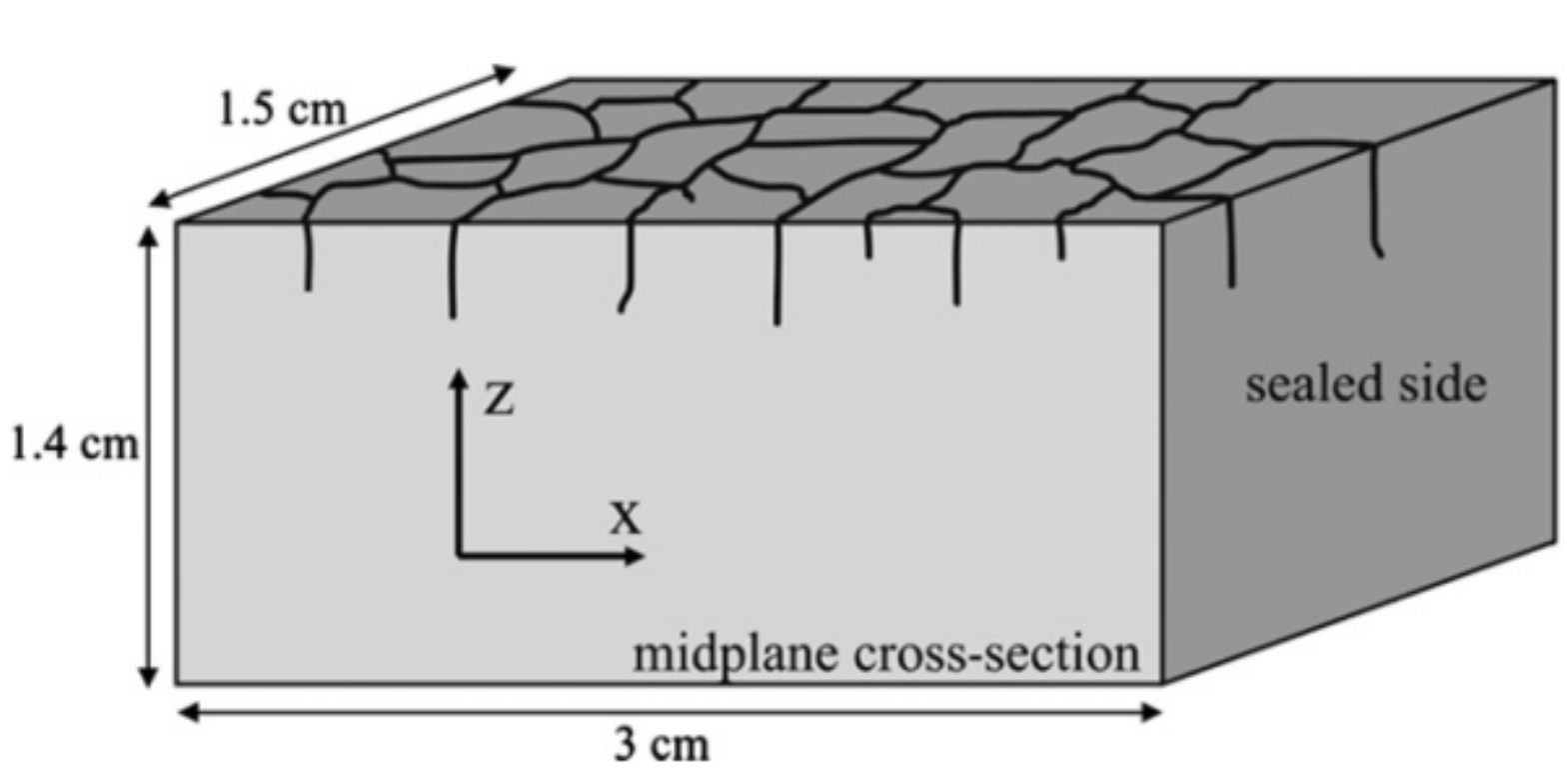} }
  \caption{Geometry of thickest \textit{ss}-sample with desiccation crack-pattern cut along its midplane.}
\end{figure}

Desiccation crack depths were measured on sample cross-sections. The impregnated samples were embedded in an Araldite adhesive, cut along their midplanes (Fig. 2), and photographed in fluorescent light. The section crack-patterns were analysed as follows: First, the 3-cm sample images were all rescaled to the same resolution (500 pixels/cm). Then, the cracks were manually traced and processed to obtain binary images of skeletonised crack-patterns. The crack length was measured in Matlab as the length of a line going through the middle points of crack pixels. The final crack length was obtained by adding the thickness of the impregnated top layer that was removed before sectioning. Note that desiccation cracks are roughly perpendicular to the drying surface and that crack depth and crack length are similar measures. In this study crack lengths were measured from low-resolution crack traces, and thus are lower than (but linearly proportional to) true crack lengths that can be measured from e.g., high magnification SEM-images.
\subsection{FEM analysis of drying process}
Drying induced moisture distributions/gradients were calculated for selected samples by finite element modelling in Abaqus. Calculations are based on the analogy of moisture diffusion to heat transfer with a convective boundary layer [30]. Fick's second law for moisture diffusion can be written as: $C/t = D(C)\nabla^2C$, with moisture content $C$ and the moisture diffusivity $D(C)$. The moisture dependence of $D(C)$ is expressed by $D(C) = (a + b(1-2exp[10d(C-1)]))$ following [31]. The moisture diffusion at the drying surface is considered as a convective boundary condition and reads: $D(C)\cdot\partial C/\partial n = h_m (C_s-C_e)$ [30], where $\partial C/\partial n$ is the moisture gradient at the surface with unit normal $n$; $h_m$ is the convective hygral transfer coefficient (or film coefficient); $C_s$ is the moisture content at the drying surface; $C_e$ is the moisture content of the far-field environment at 26\% RH. The transfer coefficient $h_m$ under ambient conditions was taken to be 1 and 10 cm/day for the \textit{non-shock} and \textit{shock} experiments, respectively, based on values given for drying concrete [15,30]. The fitting parameters $a$, $b$, and $d$ in the diffusion coefficient equation ($a$ = 0.1; $b$ = 0.4; and $d$ = 4) were obtained by assuming $h_m$ = 10 cm/day and fitting the parameters in $D(C)$ to numerically obtain the measured total moisture loss of the 14 mm thick \textit{shock-ss} sample drying for 2 months.
\section{Results}
\subsection{Moisture loss}
The total moisture loss of the samples was recorded by measuring their weights before and after the experiment (at 26\% RH and 20 $^\circ$C). During the first 5 h of drying there was a significant effect of drying method, but a small effect of sample thickness on moisture loss (Fig. 3). Samples drying in the ESEM lost about 1.4 times more water than the \textit{shock-ss} samples in the climate chamber. The \textit{shock-ds}-samples lost about 1.4 times more water than the \textit{non-shock-ds} samples. The \textit{ds}-samples lost about twice as much water as the \textit{ss}-samples. As in a previous study [29], no significant effect of desiccation microcracking on the drying rate has been observed in this study.
\begin{figure}[htb]
  \centering{ \includegraphics[width=10.cm]{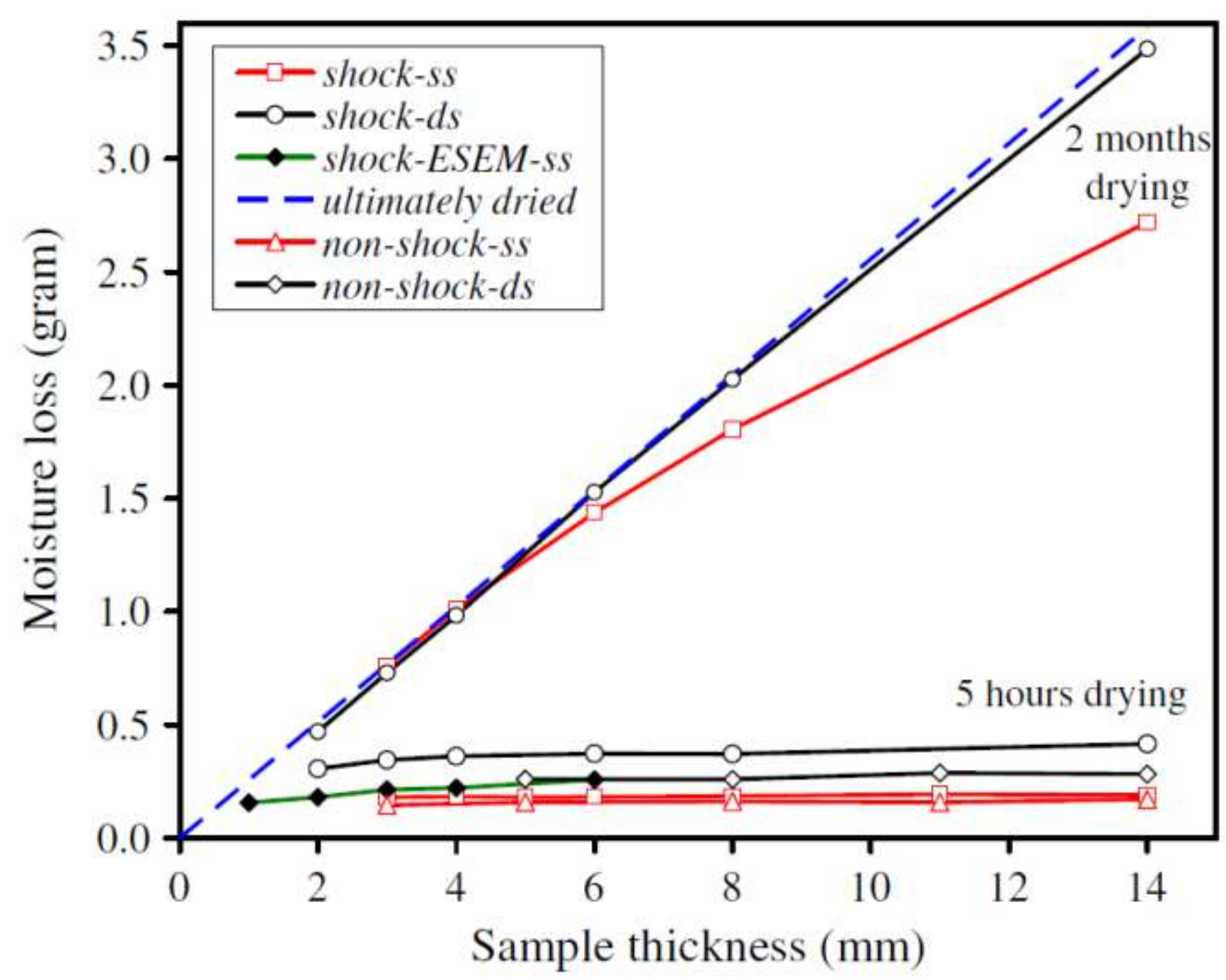} }
  \caption{Total moisture loss as function of sample thickness, drying time and drying method. Each point represents the average of three or six samples.}
\end{figure}

In the ultimate state of drying, the moisture loss should be a simple linear function of the sample volume. The dashed line in Fig. 3 is a linear fit through all the data points for \textit{shock-ds} samples with a thickness $\leq$ 8 mm. All \textit{ds}-samples, with the exception of the 14 mm ones, appear to have reached the ultimate state of drying after 2 months, whereas for the \textit{ss}-samples this was only the case for those thinner than 6 mm. The \textit{non-shock} samples showed similar moisture loss values to the \textit{shock} samples after 2 months drying (not plotted in Fig. 3).

The samples lost approximately 60\% of the free water content by drying to the ultimate state. The remaining 40\% is absorbed to the internal pore surface of the material in equilibrium with 26\%RH air at 20 $^\circ$C. Approximately 25\% of the initial water content in the mixture became chemically-bound by cement hydration.
\subsection{Crack patterns on the drying surface}
Typical desiccation crack-patterns as observed on the drying surface of the 3 x 3 cm samples are shown in Fig. 4. On the photographs the cracks appear much wider than they are in reality because the fluorescent resin penetrated part of the material adjacent cracks, and pictures were taken using long exposure times. There exists a critical sample thickness for all drying methods below which samples remained crack-free upon drying. This thickness was around 2 mm for \textit{ds-} and \textit{ESEM-ss} samples, and 3-4 mm for \textit{ss}-samples. Samples that dried in the ESEM showed much higher surface crack densities than those from the climate chamber at ambient pressure (Fig. 4). The pattern geometry in \textit{ESEM}-samples is characterised by many 120$^\circ$ crack junction angles and more dead-end cracks (Fig. 4d). Drying in the climate chamber produced more hierarchical fracture patterns with dominant 90$^\circ$/180$^\circ$ crack junctions [32]. Further effects of sample thickness and drying method will be described on the basis of sample section measurements.
\begin{figure}[htb]
  \centering{ \includegraphics[width=16.cm]{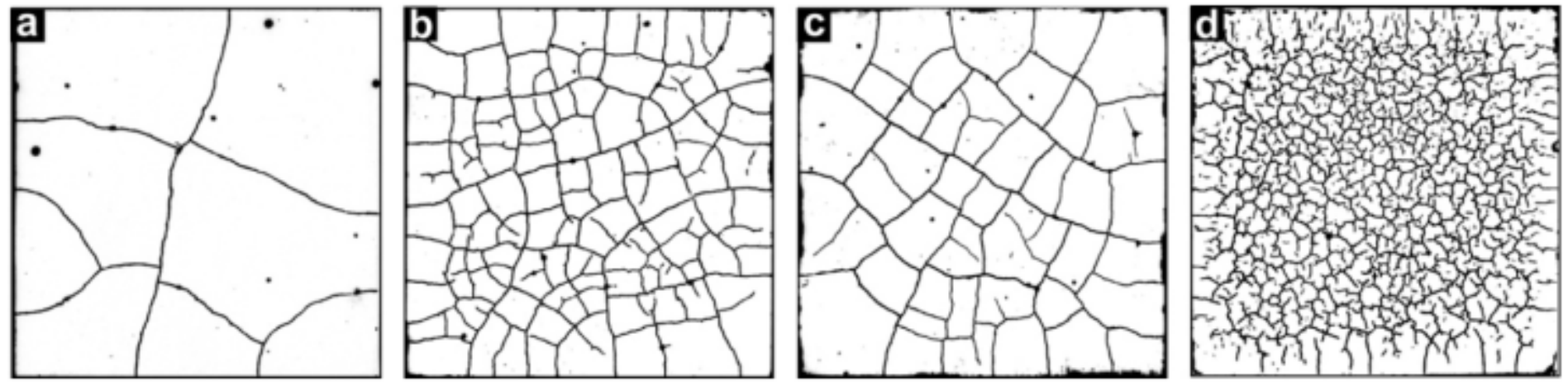} }
  \caption{Crack-pattern examples on the drying surface of 3 x 3 cm samples (colour-inverted images). (a) Top surface of \textit{non-shock-ds} sample after 5 h drying (d = 8 mm); (b.) top surface of \textit{shock-ds} sample after 5 h drying (d = 8 mm); (c) top surface of \textit{shock-ds} sample after 2 months drying (d = 8 mm); and (d) \textit{ESEM}-sample after 5 h drying (d = 6 mm).}
\end{figure}

Desiccation crack widths (on the drying surface) were measured \textit{in situ} in the Environmental SEM and \textit{ex-situ} in the impregnated samples that dried in the climate chamber. Desiccation cracks in the \textit{ESEM}-samples were observed as soon as imaging was possible (4-5 min after closing the ESEM-chamber). The first observed crack widths are typically around 1 $\mu$m in samples with thicknesses of 3, 4, and 6 mm. The width evolution in the first 5 h of drying was different for different cracks in the same sample (Fig. 5). Some cracks continuously opened and reached a width of 4 $\mu$m at the end of the 5 h experiment (Fig. 5a-c). Other cracks reached their maximum crack width after approximately 1 h drying, and then started to close again (Fig. 5d-e). A few cracks continuously decreased in width during the 5 h experiment, possibly as a consequence of opening of other cracks.
\begin{figure}[htb]
  \centering{ \includegraphics[width=8.cm]{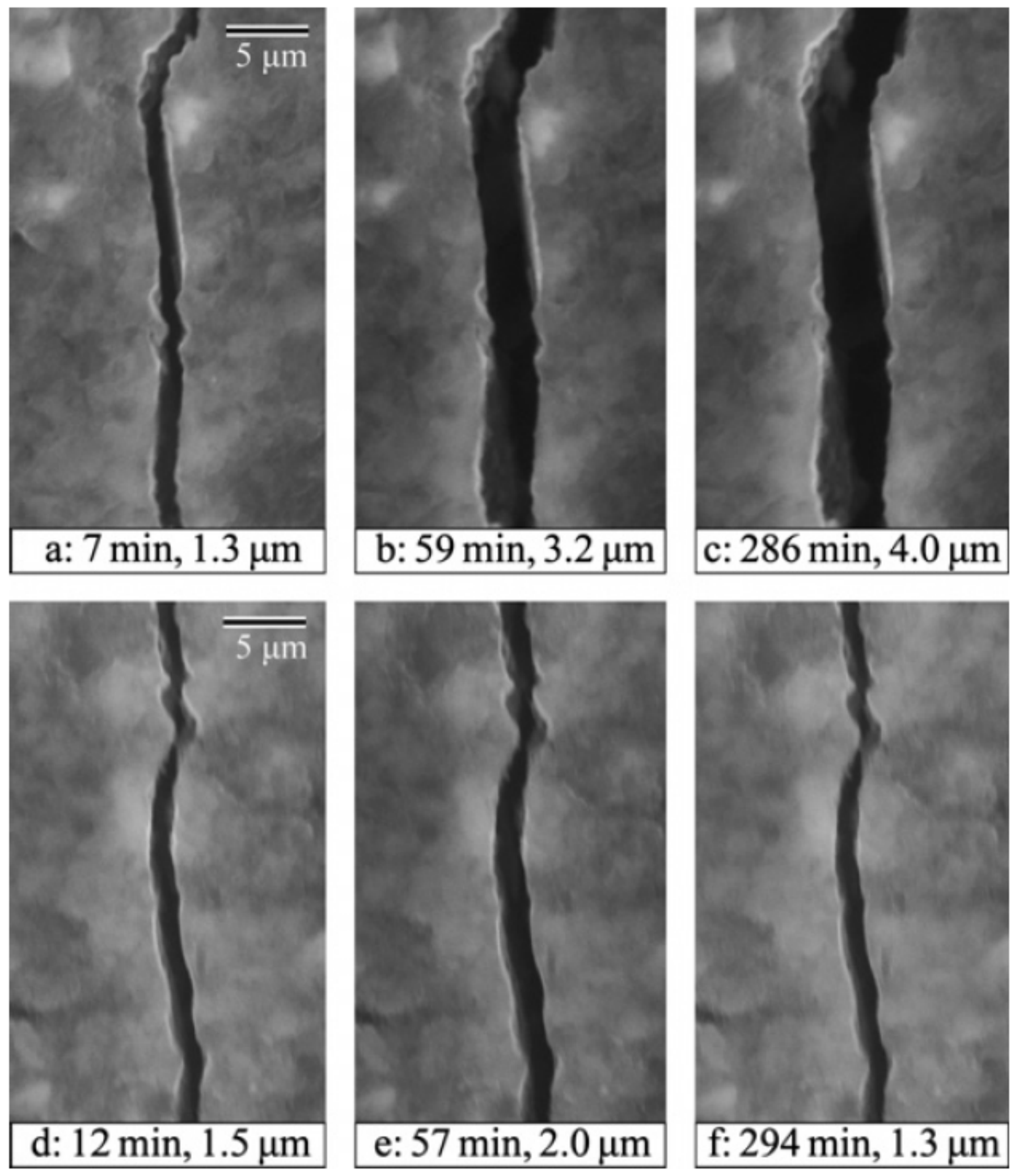} }
  \caption{Crack width evolution of two desiccation cracks (a–c) and (d–f) in the same 6 mm thick sample (ESEM images).}
\end{figure}

The desiccation crack widths in samples that dried in the climate chamber were measured from the impregnated samples using SEM. The maximum crack width after 5 h drying in the \textit{shock-ss} and \textit{shock-ds} series was 6 $\mu$m in 14 mm thick samples. After 2 months drying the maximum crack width in 14 mm thick samples was 7 $\mu$m in \textit{shock-ss} samples and 9 $\mu$m for \textit{shock-ds} samples, and most of the visible cracks had widths well above 1 $\mu$m. Also in thinner samples, crack widths after 2 months were not smaller than those after 5 h drying.
\subsection{Crack patterns on sample sections}
Typical desiccation crack-patterns on sample cross-sections are shown in Fig. 6. A general trend for all drying methods is that the average and maximum crack depth and the variation in crack depth increased with sample thickness. The \textit{ESEM}-samples show a clear bimodal distribution of crack lengths for sample thicknesses $\leq$4mm(Fig. 7a). In \textit{shock-ds} samples the distribution widens
and becomes more asymmetrical with increasing sample thickness, but no clear bimodal behaviour is observed in these samples (Fig. 7b). Only cracks in the top halves of the \textit{shock-ds} samples are plotted, because the depth of cracks in bottom halves was generally lower (Fig. 8). The maximum crack depths in \textit{shock-ss} samples ranged between 15\% and 20\% and in \textit{shock-ds} samples between 20\% and 35\% of the sample thickness (Fig. 8).
\begin{figure}[htb]
  \centering{ \includegraphics[width=16.cm]{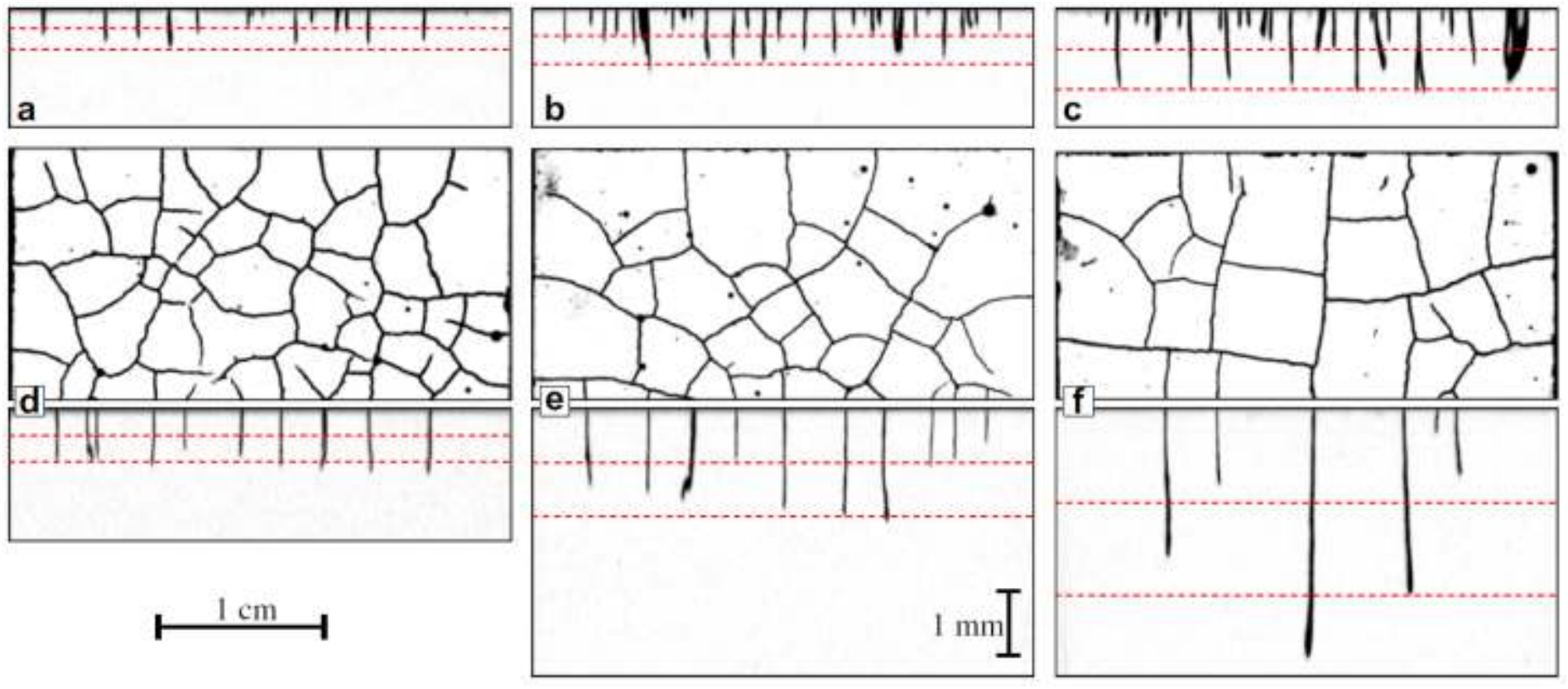} }
  \caption{(a–c) Crack-patterns after 5 h drying on cross-sections of \textit{ESEM}-samples with thickness of 3 mm (a), 4 mm (b), and 6 mm (c). (d–f) Crack-patterns after 2 months drying on top surfaces and corresponding midplane sections of \textit{shock-ds} samples with thickness of 4 mm (d), 8 mm (e), and 14 mm (f). The two dashed (red) lines in each image indicate depths levels of 10\% and 20\% of the sample thickness. All sections images are stretched in the vertical direction by a factor four.}
\end{figure}

\begin{figure}[htb]
  \centering{ \includegraphics[width=14.cm]{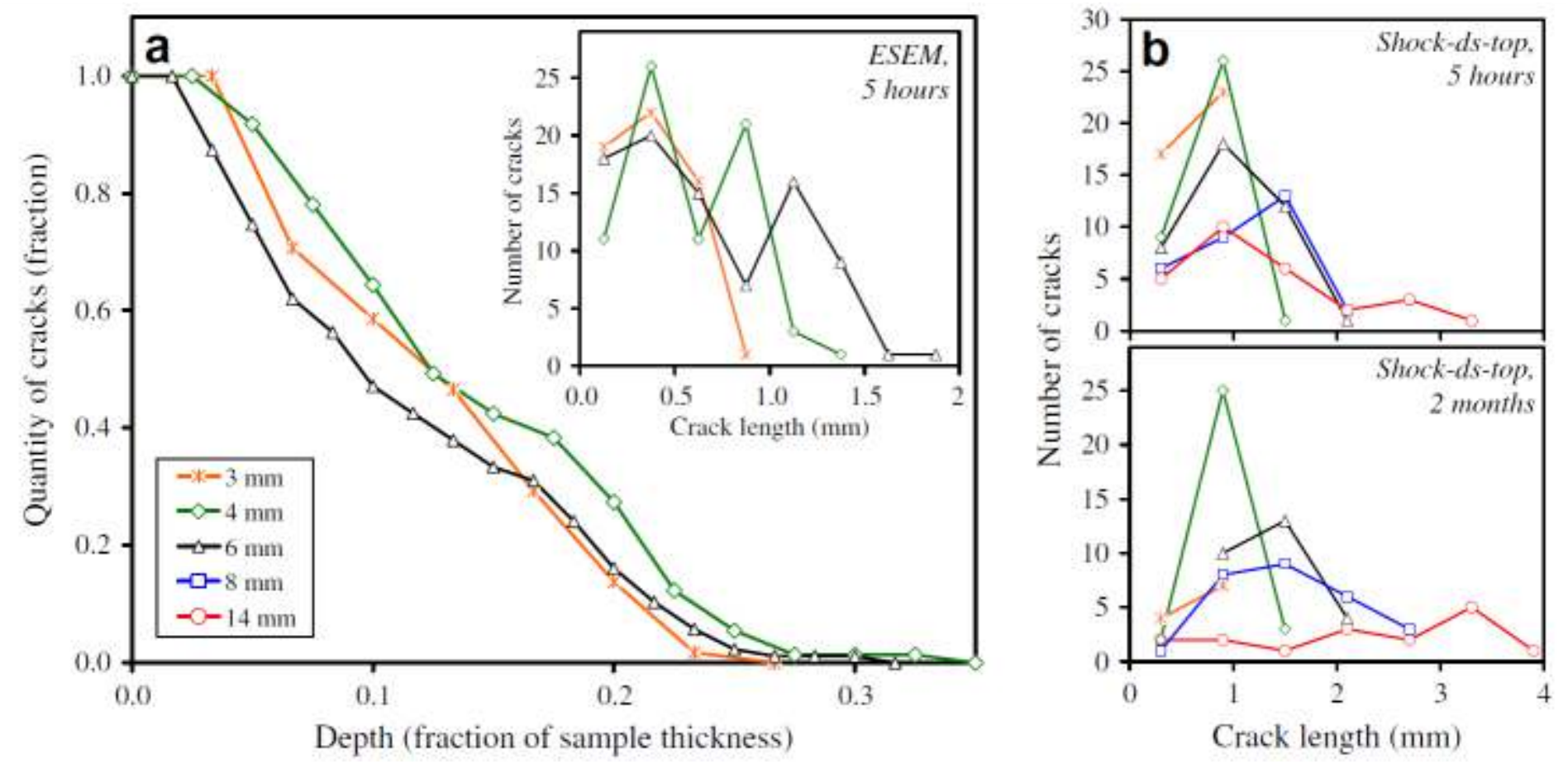} }
  \caption{(a) \textit{ESEM}-samples. Main graph shows the number of cracks as function of distance from the drying surface and the inset shows the corresponding crack length distributions; (b) Crack length distributions of \textit{shock-ds} samples (top surfaces only). The graphs show totals of three samples from each thickness category.}
\end{figure}

\begin{figure}[htb]
  \centering{ \includegraphics[width=10.cm]{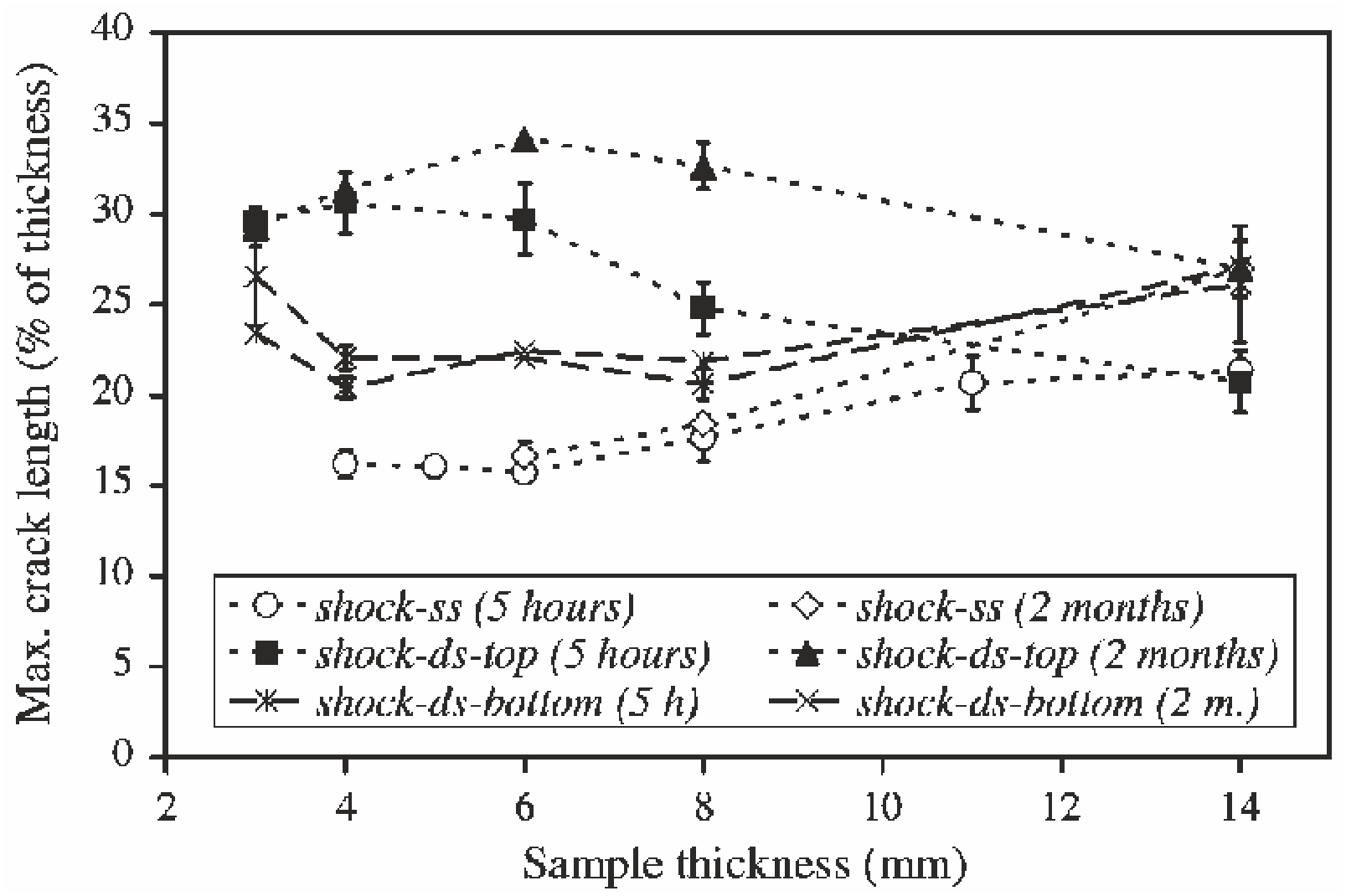} }
  \caption{Maximum crack depth as proportion of sample thickness for \textit{shock-ss} and \textit{shock-ds} samples. Data points are averages of the two longest cracks from three samples and error bars show the maximum and minimum values.}
\end{figure}

Total cumulative crack lengths on the 3 cm long sections are plotted in Fig. 9. These total crack lengths or section crack densities are proportional to 3D crack densities by a factor 1/2$\pi$ [28]. As mentioned before, there exists a critical sample thickness for each drying method below which the samples remained crack-free upon drying. Above the critical thickness, the section crack densities increase up to a sample thickness of 14 mm at which the curves for \textit{shock-ss} and \textit{shock-ds} samples merge. Drying rate had a significant effect on section crack densities. Samples that dried in the ESEM reached a crack density about twice as high as the most severely dried samples in the climate chamber after 5 h drying (Fig. 9a). Decreasing the drying rate in the first 5 h (in \textit{non-shock} samples), by placing them in a cylinder, had a strong effect on the crack density after 5 h (Fig. 9a). This effect disappeared in subsequent drying for 2 months outside the cylinders (Fig. 9b), meaning that cracking in these samples was simply delayed.
\begin{figure}[htb]
  \centering{ \includegraphics[width=8.cm]{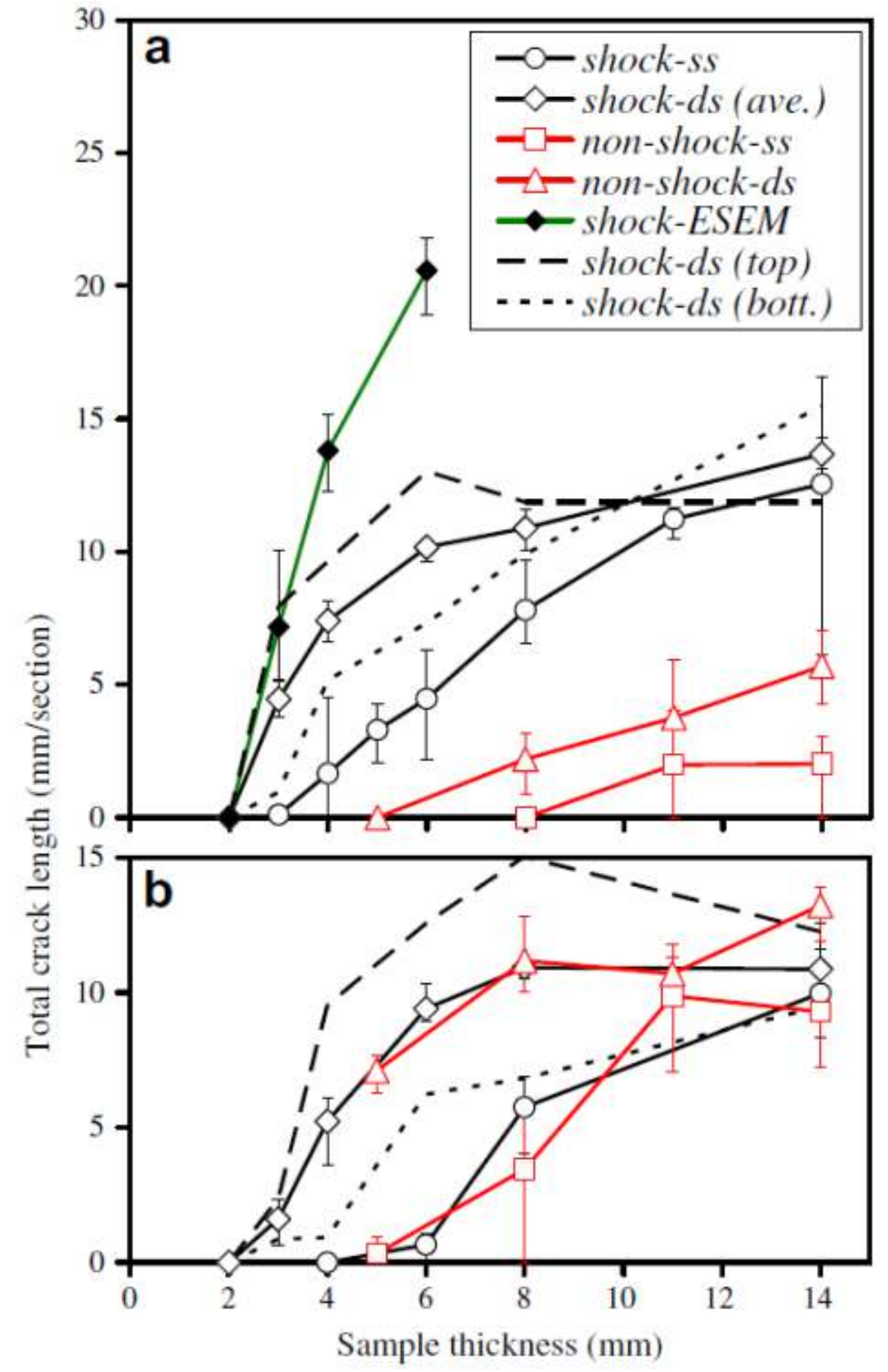} }
  \caption{Average cumulative crack length on sample cross-section (in mm/section) measured after (a) 5 h drying, and (b) 2 months drying. For \textit{ds}-samples the values are divided by two to make them comparable to \textit{ss}-samples. Error bars show maximum and minimum values of three samples.}
\end{figure}

Double-sided drying led consistently to higher section crack densities (per drying surface) compared to single-sided drying (Fig. 9). Compaction of the mixtures resulted in a density gradient in the samples (Section 2.1) and this caused the crack densities in top halves to be significantly higher than in bottom halves of \textit{ds}-samples (dashed lines in Fig. 9). Note that crack densities in bottom
halves of \textit{ds}-samples were always higher than in the corresponding \textit{ss}-samples that also dried from the bottom surface.

The section crack densities after 2 months drying are lower (up to 25\%) for the \textit{shock-ss} and \textit{shock-ds} than those after 5 h drying. The width of cracks on the drying surface was larger after 2 months drying (see Section 3.2), so it is unlikely that this result was due to complete closure of cracks in the ultimate state of drying. We explain this difference by variation between series of samples that dried not simultaneously and came from different cement paste batches.
\section{Discussion}
\subsection{Effect of drying rate}
Desiccation cracking in unconfined hardened cement paste samples is caused by the initial development of moisture and contraction gradients. In the studied samples (up to 14 mm thickness) most of the desiccation cracks formed in the first 5 h of drying (Fig. 9). Using the acoustic emission technique it has been shown before that most desiccation cracks in cement paste appear in the first hour of drying under similar drying conditions [33], even in unconfined samples with a thickness of 40 mm [34]. In order to better explain the experimental results, the development of the moisture gradient in the first 5 h of drying was simulated using finite element modelling (see Section 2.4.). This was done for the 6 mm thick samples that experienced shock drying in the climate chamber and in the ESEM-chamber. Drying of \textit{non-shock} samples was also simulated and compared to interpolated experimental results obtained from 5 and 8 mm thick samples. In all simulations and experiments the conditions of the far-field were held the same at 26\% RH and 20 $^\circ$C. The only model parameters that were varied to simulate the experimental results were the diffusion coefficient $D(C)$ and film or convective hygral transfer coefficient $h_m$. The simulated moisture distributions are shown in Fig. 10 with free moisture content given in grams per volume unit of cement paste. The total free (non-chemically bound) moisture content was measured by completely drying samples at a 28 day age in an oven at 105 $^\circ$C.
\begin{figure}[htb]
  \centering{ \includegraphics[width=14.cm]{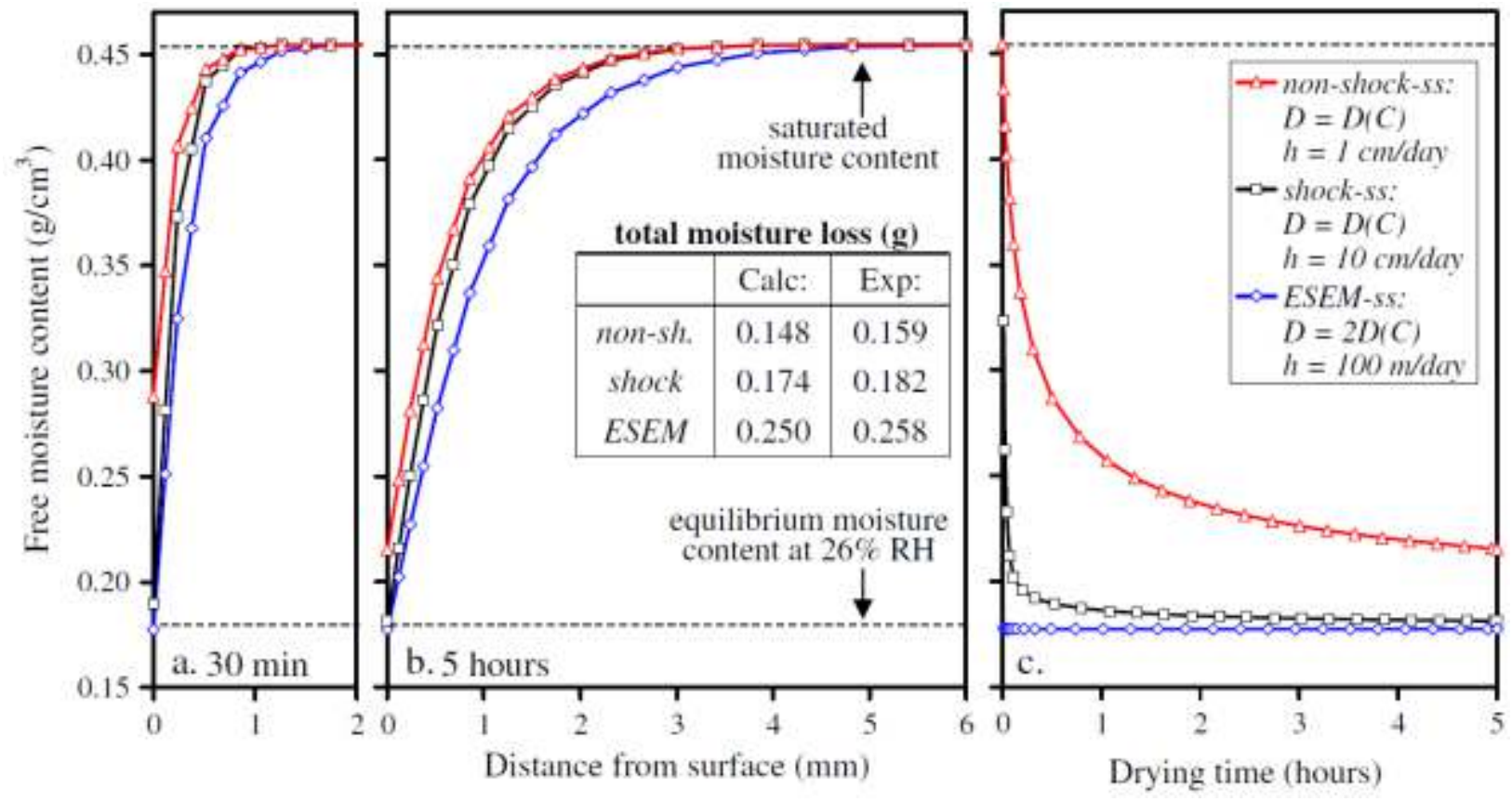} }
  \caption{Calculated moisture gradients after (a) 30 min and (b) 5 h of drying in 30 x 30 x 6 mm samples. Table inset shows calculated and experimentally measured total moisture loss after 5 h of drying. (c) Calculated moisture content at the drying surface vs. drying time ($D$ is diffusion coefficient; $h$ is film coefficient).}
\end{figure}

The total moisture losses of the \textit{shock} and \textit{non-shock} samples after 5 h drying are reasonably well predicted if the film coefficient in the \textit{non-shock} case is chosen to be a factor 10 smaller (Fig. 10). The total moisture loss of the \textit{ESEM}-samples could not predicted by just increasing the film coefficient. For example, for an arbitrarily high film coefficient of 100 m/day, the diffusion coefficient $D(C)$ needed to be multiplied by a factor 2 to simulate the observed moisture loss (Fig. 10). Both, the diffusion and film coefficient in ESEM-drying are expected to be higher than in drying under ambient conditions. The diffusion coefficient of water vapour in the ESEM-'vacuum' (at 590 Pa) is expected to be about 170 times larger than the one at atmospheric pressure in the climate chamber ($\approx 10^5$ Pa), since the diffusion coefficient of gasses is roughly inversely proportional to the gas pressure at constant temperature [35]. In the convection-free case with equal boundary layer (film) thickness, the film coefficient for ESEM-drying would thus be 170 times larger than for drying under ambient pressure. The higher diffusion coefficient of \textit{ESEM}-samples can be explained by the gradual development of 'vacuum' inside the pores of the sample that increases the vapour transport and evaporation rate in the material compared to the ambient pressure case.

The simulations in Fig. 10 show that the film coefficient not only determines the sample drying rate, but also the magnitude of the moisture gradient, if the moisture gradient is defined as the difference
between maximum and minimum moisture content in the sample. For example, after 30 min and 5 h drying, the moisture gradient in the shock case is a factor 1.6 and 1.2 larger, respectively, than the one in the \textit{non-shock} case (Fig. 10c). The \textit{ESEM}-samples are predicted to reach the maximum moisture gradient in the first minutes of drying. In the first hour of drying, when most desiccation
cracks develop [33,34], and especially at the onset of drying, large differences in moisture gradient magnitude probably existed in the \textit{non-shock}, \textit{shock}, and \textit{ESEM}-samples. This qualitatively explains the differences in the degree of desiccation cracking in these samples after 5 h drying (Fig. 9a). The quantitative relationship between moisture gradient magnitude and degree of
cracking needs to be studied in hygro-mechanical models including formulations for the relationships between (i) moisture gradient and shrinkage gradient; (ii) shrinkage gradient and stress gradient; (iii) stress and creep; and (iv) stress and cracking. In this paper we show that choice of the convective boundary condition in hygro-mechanical models is expected to be important. In this study, a 10 times larger film coefficient seems to have caused a 2-6 times larger desiccation crack density at ambient conditions with equal relative humidity (Fig. 9a).
\subsection{Effect of sample warping and thickness}
The degree of desiccation cracking in unconfined samples does not only depend on how the moisture/contraction gradient develops, it also depends on the capability of the samples to respond to the contraction gradient by warping. Warping is macroscopic curving of samples that dry single-sided and it reduces drying-induced stresses compared to equal samples that are not allowed to warp. While it is likely that the maximum sample warping occurs much later than crack-initiation/formation, we do suspect an initial warping effect on desiccation cracking in this study. This effect emerges when desiccation cracking in single-sided and double sided drying samples are compared. Thin samples drying from two opposed sides (\textit{shock-ds} and \textit{non-shock-ds}) showed significantly more and deeper cracking than corresponding \textit{ss}-samples (Figs. 8 and 9) and we explain this result by the 'obstruction' of initial warping in \textit{ds}-samples.

Three possible reasons for the observed effect of sample thickness on desiccation cracking are as follows: (i) An effect of sample thickness on the initial moisture gradient. This effect is important in the range of sample thicknesses, in which the initial moisture gradient required to initiate cracking, stretches over a depth that approaches the sample thickness. This could be the main reason why there exists a critical sample thickness for all drying methods below which no desiccation cracking occurs (Fig. 9). (ii) Thinner \textit{ss}-samples probably undergo larger initial warping and thus developed lower surface stresses and therefore less cracking. This warping effect is probably responsible for the critical crack-free sample thickness of \textit{ds}-samples being lower than the one for the \textit{ss} samples (Fig. 9). (iii) The amount of post-critical crack growth, as will be explained in Section 4.3, increases with sample thickness.
\subsection{Analysis of crack depth distributions}
A distinctive feature of shrinkage crack-patterns due to global contraction gradients is the variation of crack depths, i.e., undulations of the front formed by the crack tips [19-27]. At the onset of
drying or cooling, an initial system of parallel cracks with a specific spacing and depth is formed. The growth of equally long cracks continuous until a critical point is reached where the stress field
of adjacent cracks starts to significantly interact. From this point onwards only every second crack continues to grow since intermediate cracks become 'shielded' by their neighbours, and they eventually
close [19-25]. This process may repeat itself a number of times and results, in the theoretical 2-dimensional case, in a crack-pattern as shown in Fig. 11. Note that Fig. 11 depicts a simplified geometrical representation of the above-mentioned crack spacing doubling models, and is not based on hygro-mechanical calculations. The maximum crack depth in unconfined samples has been reported to depend primarily on the ratio between the free shrinkage strain es and the failure strain in tension $\varepsilon_f$ [17].
\begin{figure}[htb]
  \centering{ \includegraphics[width=8.cm]{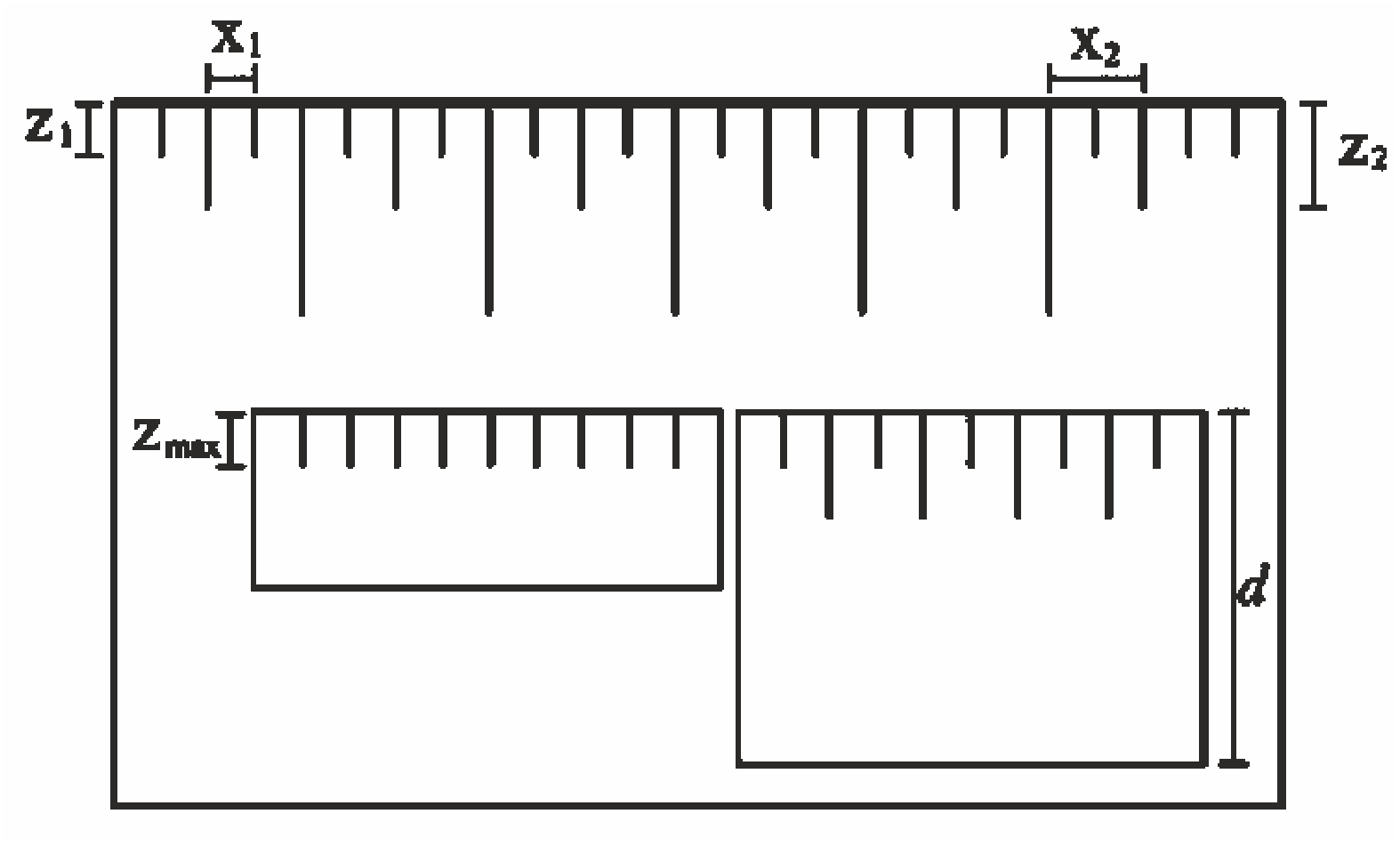} }
  \caption{Simplified 2D crack spacing doubling model for desiccation cracking of stiff solids. Maximum crack depth ($Z_{max}$) is assumed to be 30\% of sample thickness (conf. Fig. 8 for \textit{ds}-samples). Insets show expected effects of sample thickness on the crack-patterns.}
\end{figure}

The crack depth distribution diagram of the \textit{ESEM}-samples clearly shows two peaks for samples $\geq$4mm (Fig. 7a), and this is consistent with the model described above. The crack-patterns in Fig. 6a-c are, however, much less regular than the theoretical 2D model because they are sections of 3D crack-patterns. Samples dried in the climate chamber showed an increased variation in crack depth with sample thickness, but in no single sample a clear crack doubling process can be observed (Fig. 6a-f). This probably is due to the small size of these samples leading to a low number of cracks. Also sectioning effects and natural (Gaussian) variation in crack populations, contribute to crack-patterns in which a doubling process may not be directly obvious.

It is predicted that short intermediate cracks close as a consequence of opening of the adjacent deeper cracks [20,21]. In the ultimate state of drying, the shrinkage gradients disappear and all cracks are expected to close [9]. In the ESEM experiments some cracks decreased in width at the start of drying (Fig. 5d-f), and this is consistent with theoretical predictions. However, the crack widths observed after 2 months drying were not smaller than after 5 h drying, even in samples that dried completely. This could have been caused by permanent, non-uniform creep deformation of the samples under a contraction gradient.

We will now estimate geometrical crack-pattern characteristics from the average behaviour of many samples. These values are helpful for tuning the models in the above-mentioned references to hardened cement paste. For the \textit{ESEM}-samples the position of the first peak in Fig. 7a occurs at a crack length of 0.38 mmfor samples with 3, 4 or 6 mm thickness. This shows that the depth of the initial set of cracks was unaffected by sample thickness. The average crack spacing is 1.96, 1.26, and 1.23 mm for \textit{ESEM}-samples with 3, 4, and 6 mm thickness, respectively. The significantly higher value for the 3 mm samples is probably due to their larger warping potential as discussed earlier. The depth to spacing ratio of the initial set of cracks ($Z_1/X_1$) for the 4 and 6 mm \textit{ESEM}-samples are
thus 0.30 and 0.31, respectively. The position of the second peak occurs at a mean crack depth of 1.13 mm in 6 mm samples. By assuming that the spacing doubles for the second set of cracks, a lower bound estimate of depth to spacing ratio ($Z_2/X_2$) of 0.46 for \textit{ESEM}-samples is obtained. The same estimate is made for crack-patterns in \textit{shock-ds} samples that do not warp and show a relative high number of cracks. Because of the gradient density in the samples only the crack-patterns in the sample top halves are considered. The depth of the initial crack set ($Z_1$) was measured in 4 mm thick samples. These are the thickest samples with a symmetrical (Gaussian) distribution of crack depths. In thicker samples the crack depth distributions become wider and more asymmetrical due to bifurcation of the crack patterns (Fig. 7b). The mean crack depth in these samples was 0.79 and 0.95 mm after 5 h and 2 months drying, respectively. With average crack spacings of 2.61 and 3.21 mm, respectively, an average ratio $Z_1/X_1$ of 0.30 for all 4 mm thick \textit{shock-ds} samples is obtained.

It is not possible to give a reliable lower bound estimate of the ratio $Z_2/X_2$ for \textit{shock-ds} samples because of a too low number of second generation cracks in these samples. Studying wider and thicker (\textit{ss} or \textit{ds}) samples will probably improve statistics and yield crack depth distribution diagrams with clearer, multiple peaks. However, the formation of residual thermal stresses by the cement hydration process becomes more important in larger specimens and these stresses may affect crack depth distributions, especially in young samples [11]. 
\subsection{Crack-free drying}
A motivation for this study was to find a method to reduce or avoid desiccation cracking during drying of small hardened cement paste samples. Dried, crack-free samples are needed to measure the effect of moisture content or relative humidity on the mechanical properties, such as strength, elasticity, creep and pure drying shrinkage of hardened cement paste. Samples up to 3 mm thickness can be dried crack-free down to 26\% relative humidity (under atmospheric pressure) if they are dried from a single side. The critical crack-free sample thickness can probably be increased by reducing the drying rate. This can be achieved by reducing the velocity of air flow past the sample surface or by applying a stepwise drying procedure. However, avoiding cracking in thicker samples by slower drying may not be practical because of drying times in the order of many months or even years [10]. Reducing just the initial drying rate to avoid the development of large contraction gradients at the onset of drying was not a successful method in this study: i.e., the \textit{non-shock} samples showed equal crack densities as the \textit{shock} samples in the ultimate state of drying. Perhaps this method works if the initial drying rate is not reduced for 5 h but rather for a number of days. Finally, a way to obtain crack-free samples is to dry the samples under compressive load as proposed or observed before [10,11,18].
\section{Conclusions}
\begin{itemize}
\item The depth of desiccation cracks scaled roughly with sample thickness in unconfined samples of hardened cement paste. For sample thicknesses < 14 mm, single-sided drying lead to significantly lower crack depths (15-20\% of thickness) than double-sided drying (20-35\% of thickness). The difference is explained by the capability of single-sided samples to respond to contraction gradients by warping.
\item Drying rate and single-sided vs. double-sided drying had an important effect on the critical crack-free sample thickness. Whatever the drying method, cement paste samples thinner than 2 mm are unlikely to crack upon shock drying at relative humidities above 25\%.
\item Desiccation cracking in unconfined hardened cement paste samples occurs at the onset of drying. Hardened cement paste samples taken out of curing water in a dry room will very quickly develop desiccation microcracks.
\item Crack spacing doubling models seem to apply for desiccation cracking in cement paste. The ratio between crack depth and crack spacing for first set cracks was 0.30, and for second set cracks at least 0.46.
\end{itemize}
\section*{References}
\begin{enumerate}
\item Groisman A, Kaplan E. An experimental study of cracking induced by desiccation. Europhys Lett 1994;25:15-420.
\item Colina H, Acker P. Drying cracks: kinematics and scale laws. Mater Struct 2000;33:101-7.
\item Shorlin KA, de Bruyn JR, Graham M, Morris SW. Development and geometry of isotropic and directional shrinkage-crack patterns. Phys Rev E 2000;61:6950-7.
\item Bohn S, Pauchard L, Couder Y. Hierarchical crack pattern as formed by successive domain divisions I. Temporal and geometrical hierarchy. Phys Rev E 2005;71:046214.
\item Bentur A. Early age cracking tests. In: Bentur A, editor. Early age cracking in cementitious systems. Report of RILEM Technical Committee TC 181-EAS, 2002. p. 241-55.
\item Weiss WJ, Shah SP. Restrained shrinkage cracking: the role of shrinkage reducing admixtures and specimen geometry. Mater Struct 2008;35:85-91.
\item Hossain AB, Weiss J. The role of specimen geometry and boundary conditions on stress development and cracking behaviour in the restrained ring test. Cem Concr Res 2006;36:189-99.
\item Loser R, Leemann A. Shrinkage and restrained shrinkage cracking of selfcompacting concrete compared to conventionally vibrated concrete. Mater Struct 2009;42:71-82.
\item Hwang CL, Young JF. Drying shrinkage of Portland cement pastes I. Microcracking during drying. Cem Concr Res 1984;14:585-94.
\item Bazant ZP, Raftshol WJ. Effect of cracking in drying and shrinkage specimens. Cem Concr Res 1982;12:209-26.
\item Bisschop J. Size and boundary effects on desiccation cracking in hardened cement paste. Int J Fracture 2008;154:211-24.
\item Lindquist WD, Darwin D, Browning J, Miller GG. Effect of cracking on chloride content in concrete bridge decks. ACI Mater J 2006;103:467-73.
\item Ismail M, Toumi A, François R, Gagné R. Effect of crack opening on the local diffusion of chloride in cracked mortar samples. Cem Concr Res 2008;38:1106-11.
\item Northcott GDS. Concrete pavements. In: Mays GC, editor. Durability of concrete structures - investigation repair protection. London: Taylor and Francis; 1992. p. 226-47.
\item Foos S, Mechtcherine V, Mueller HS. Deformation behaviour of concrete highway pavements. In: Hendriks, Rots, editors. Finite elements in civil engineering applications. The Netherlands: Swets and Zeitlinger B.V., Lisse; 2002. p. 123-7.
\item Valenza II JJ, Scherer GW. Mechanism for salt scaling. J Am Ceram Soc 2006;89:1161-79.
\item Thelandersson S, Martensson A, Dahlblom O. Tension softening and cracking in drying concrete. Mater Struct 1988;21:416-24.
\item de Sa C, Benboudjema F, Thiery M, Sicard J. Analysis of microcracking induced by differential drying shrinkage. Cem Concr Comp 2008;30:947-56.
\item Granger L, Torrenti JM, Acker P. Thoughts about drying shrinkage: experimental results and quantification of structural drying creep. Mater Struct 1997;30:588-98.
\item Nemat-Nasser S, Keer LM, Parihar KS. Unstable growth of thermally induced interacting cracks in brittle solids. Int J Sol Struc 1978;14:409-29.
\item Bazant ZP, Ohtsubo H, Aoh K. Stability and post-critical growth of a system of cooling or shrinkage cracks. Int J Fracture 1979;15:443-56.
\item Li YN, Hong AP, Baz?ant ZP. Initiation of parallel cracks from surface of elastic half-plane. Int J Fracture 1995;69:357-69.
\item Jagla EA. Stable propagation of an ordered array of cracks during directional drying. Phys Rev E 2002;65:046147.
\item Bahr HA, Weiss HJ, Maschke HG, Meissner F. Multiple crack propagation in a strip caused by thermal shock. Theor Appl Fract Mech 1988;10:219-26.
\item Jenkins DR. Optimal spacing and penetration of cracks in a shrinking slab. Phys Rev E 2005;71:056117.
\item Bahr HA, Fisher G, Weiss HJ. Thermal-shock crack pattern explained by single and multiple crack propagation. J Mater Sci 1986;21:2716-20.
\item Yuan C, Vandeperre LJ, Stearn RJ, Clegg WJ. The effect of porosity in thermal shock. J Mater Sci 2008;43:4099-106.
\item Bisschop J. Drying shrinkage microcracking in cement-based materials. PhDthesis Delft University of Technology, the Netherlands: Delft University Press; 2002 (ISBN 90-407-2341-9).
\item Bisschop J, Van Mier JGM. Effect of aggregates and microcracks on the drying rate of cementitious composites. Cem Concr Res 2008;38:1190-6.
\item Baluch MH, Rahman MK, Mahmoud IA. Calculating drying shrinkage stresses. Concr Int 2008;30:37-41.
\item Xi Y, Baz?ant ZP, Molina L, Jennings HM. Moisture diffusion in cementitious materials - Moisture capacity and diffusivity. Adv Cem Based Mater 1994;1:258-66.
\item Toga KB, Erdem Alaca B. Junction formation during desiccation cracking. Phys Rev E 2006;74:021405.
\item Nadeau JS, Bennet R, Mindess S. Acoustic emission in the drying of hardened cement paste and mortar. J Am Ceram Soc 1981;64:410-5.
\item Shiotani T, Bisschop J, Van Mier JGM. Temporal and spatial development of drying shrinkage cracking in cement-based materials. Eng Fract Mech 2003;70:1509-25.
\item Cussler EL. Diffusion - mass transfer in fluid systems. 2nd ed. Cambridge, New York: Cambridge University Press; 1997. p. 109.
\end{enumerate}

\end{document}